\documentstyle[aps,epsfig]{revtex}
\begin{document}

\title{ Equation of state for nuclear matter based on density dependent effective interaction }

\author{D.N. Basu\thanks{E-mail:dnb@veccal.ernet.in}}
\address{Variable  Energy  Cyclotron  Centre,  1/AF Bidhan Nagar,
Kolkata 700 064, India}
\date{\today }
\maketitle
\begin{abstract}

      An interesting method of obtaining equation of state for nuclear matter, from a density dependent M3Y interaction, by minimizing the energy per nucleon is described. The density dependence parameters of the interaction are obtained by reproducing the saturation energy per nucleon and the saturation density of spin and isospin symmetric cold infinite nuclear matter. The nuclear matter equation of state thus obtained is then used to calculate the pressure, the energy density, the nuclear incompressibility and the velocity of sound in nuclear medium. The results obtained are in good agreement with experimental data and provide a unified description of radioactivity, scattering and nuclear matter.

\end{abstract}

\pacs{ PACS numbers: 21.65.+f, 23.60.+e, 23.70.+j, 25.70.Bc, 21.30.Fe, 24.10.Ht }   


\section{Introduction}
\label{section1}

      Nuclear matter is an idealized system of nucleons interacting without Coulomb forces which is translationally invariant with a fixed ratio of neutrons to protons. The goal of nuclear matter theory is to obtain an equation of state (EOS) for nuclear matter starting from the underlying two-body nucleon-nucleon (NN) interaction. The nuclear EOS, which is the energy per nucleon E/A = $\epsilon$ of nuclear matter as a function of nucleonic density $\rho$, can then be used to obtain the bulk properties of nuclear matter such as the nuclear incompressibility, the energy density, the pressure, the velocity of sound in nuclear medium etc. The EOS is also of fundamental importance in the theories of nucleus-nucleus collisions at energies where the nuclear incompressibility $K$ comes into play as well as in the theories of neutron stars and supernova explosions \cite{r1} and a widely used experimental method is the determination of the incompressibility from the observed giant monopole resonances (GMR) \cite{r2}. In the present work we obtain an EOS for nuclear matter using the M3Y-Reid-Elliott effective interaction supplemented by a zero range pseudo-potential representing the single-nucleon exchange term along with the density dependence which takes care of the higher order exchange effects and the Pauli blocking effects. The density dependence parameters of the interaction are obtained by reproducing the saturation energy per nucleon and the saturation density of spin and isospin symmetric cold infinite nuclear matter. One of the density dependence parameter is also utilized to provide  estimate for the nuclear mean free path.  The EOS, thus obtained, then used to calculate the pressure, the energy density, the nuclear incompressibility and the velocity of sound in nuclear matter. 

      The M3Y interaction, which was derived by fitting its matrix elements in an oscillator basis to those elements of the G-matrix obtained with the Reid-Elliott soft-core NN interaction, has a profound physical basis. The ranges of the M3Y forces were chosen to ensure a long-range tail of the one-pion exchange potential as well as a short range repulsive part simulating the exchange of heavier mesons \cite{r3}. The real part of the nuclear interaction potential obtained by folding in the density distribution functions of two interacting nuclei with the density dependent M3Y effective interaction supplemented by a zero-range pseudo-potential (DDM3Y) was shown to provide good descriptions for medium and high energy $\alpha$ and heavy ion elastic scatterings \cite{r4,r5}. The zero-range pseudo-potential represented the single-nucleon exchange term while the density dependence accounted for the higher order exchange effects and the Pauli blocking effects. Since the density dependence of the effective projectile-nucleon interaction was found to be fairly independent of the projectile \cite{r6}, as long as the projectile-nucleus interaction was amenable to a single-folding prescription, the density dependent effects on the nucleon-nucleon interaction were factorized into a target term times a projectile term and used successfully in case of nuclear scattering \cite{r7} and cluster radioactivity \cite{r8}. Present work uses the same effective interaction within single folding model description to obtain nuclear EOS and provide unified descriptions of nuclear scattering, cluster radioactivity and nuclear matter.   

\section{The density dependent realistic effective nucleon-nucleon interaction }
\label{section2}

       The general expression for the density dependent effective NN interaction potential $v(s)$ is written as \cite{r7} 

\begin{equation}
 v(s,\rho, \epsilon) = t^{M3Y}(s, \epsilon) g(\rho, \epsilon)
\label{seqn1}
\end{equation}   
\noindent
where the realistic effective M3Y interaction potential \cite{r3} supplemented by a zero range pseudopotential $t^{M3Y}$ is given by  

\begin{equation}
 t^{M3Y}(s, \epsilon) = 7999 \exp( - 4s) / (4s) - 2134 \exp( - 2.5s) / (2.5s) + J_{00}(\epsilon) \delta(s)
\label{seqn2}
\end{equation}   
\noindent
where the zero-range pseudo-potential $J_{00}(\epsilon)$ representing the single-nucleon exchange term is given by \cite{r5} 

\begin{equation}
 J_{00}(\epsilon) = -276 (1 - \alpha\epsilon) (MeV.fm^3)
\label{seqn3}
\end{equation}   
\noindent
and the density dependent part \cite{r9} has been taken to be of a general form

\begin{equation}
 g(\rho, \epsilon) = C (1 - \beta(\epsilon)\rho^n) 
\label{seqn4}
\end{equation}   
\noindent
which takes care of the higher order exchange effects and the Pauli blocking effects. This density dependence changes sign at high densities which is of crucial importance in fulfilling the saturation condition as well as giving different $K_0$ values with different values of $n$ for the nuclear EOS. The value of the parameter $n=2/3$ was originally taken by Myers in the single folding calculation \cite{r9}. In fact $n=2/3$ also has a physical meaning because then $\beta$ can be interpreted as an 'in medium' effective nucleon-nucleon interaction cross-section $\sigma_0$ while the density dependent term represents interaction probability. This value of $\beta$ along with nucleonic density of infinite nuclear matter $\rho_0$ can also provide the nuclear mean free path $\lambda=1/(\rho_0 \sigma_0)$. Moreover, it also worked well in the double folding calculations with the factorized density dependence for the high energy heavy ion scattering \cite{r7} and the cluster radioactivity \cite{r8}. 

\section{Theoretical formalism for nuclear matter calculations}
\label{section3}

      The two parameters of eqn.(4), C and $\beta$, have been fitted to reproduce the saturation conditions. It is worthwhile to mention here that due to attractive character of the M3Y forces the saturation condition for cold nuclear matter is not fulfilled. However, the realistic description of nuclear matter properties can be obtained with this density dependent M3Y effective interaction. Therefore, the density dependence parameters have been obtained by reproducing the saturation energy per nucleon and the saturation nucleonic density of the spin and isospin symmetric cold  infinite nuclear matter.   

      The energy per nucleon $\epsilon$ obtained using the effective nucleon-nucleon interaction $v(s)$ for the spin and isospin symmetric cold infinite nuclear matter is given by

\begin{equation}
 \epsilon = [3\hbar^2k_F^2/10m] + g(\rho, \epsilon)\rho J_v / 2
\label{seqn5}
\end{equation}   
\noindent
where m is the nucleonic mass equal to 938.91897 $MeV/c^2$, $k_F$ given by
\begin{equation}
 k_F^3 = 1.5\pi^2\rho,
\label{seqn6}
\end{equation}                                                                                                                                           
\noindent     
is the Fermi momentum, $\rho$ is the nucleonic density while $\rho_{0}$ being the saturation density for the spin and isospin symmetric cold infinite nuclear matter and $J_v$ represents the volume integral of the M3Y interaction supplemented by the zero-range pseudopotential having the form    
 
\begin{equation}
 J_v(\epsilon)  =   \int \int \int t^{M3Y}(s, \epsilon) d^3s = 7999 (4\pi/4^3) - 2134 (4\pi/2.5^3) + J_{00}(\epsilon) 
\label{seqn7}
\end{equation}
\noindent
The eqn.(5) can be rewritten with the help of eqn.(4) as 

\begin{equation}
 \epsilon = [3\hbar^2k_F^2/10m] + [\rho J_v C (1 - \beta\rho^n)/2]  
\label{seqn8}
\end{equation}
\noindent
and differentiated with respect to $\rho$ to yield equation  

\begin{equation}
 \partial\epsilon/\partial\rho = [\hbar^2k_F^2/5m\rho] + J_v C [1 - (n+1)\beta\rho^n] /2 
\label{seqn9}
\end{equation}
\noindent
The equilibrium density of the nuclear matter is determined from the saturation condition $\partial\epsilon/\partial\rho = 0$. Then eqn.(8) and eqn.(9) with the saturation condition can be solved simultaneously for fixed values of the saturation energy per nucleon $\epsilon_0$ and the saturation density $\rho_{0}$ of the spin and isospin symmetric cold  infinite nuclear matter, henceforth will be called the standard nuclear matter, to obtain the values of the density dependence parameters $\beta$ and C. Density dependence parameters $\beta$ and C, thus obtained, can be given by  

\begin{equation}
 \beta = [(1-p)\rho_{0}^{-n}]/[(3n+1)-(n+1)p],
\label{seqn10}
\end{equation} 
\noindent

\begin{equation}
 p = [10m\epsilon_0]/[\hbar^2k_{F_0}^2],
\label{seqn11}
\end{equation} 
\noindent
 
\begin{equation}
 k_{F_0} = [1.5\pi^2\rho_0]^{1/3},
\label{seqn12}
\end{equation} 
\noindent

\begin{equation}
 C = -[2\hbar^2k_{F_0}^2] / [5mJ_v\rho_0(1 - (n+1)\beta\rho_0^n)],
\label{seqn13}
\end{equation} 
\noindent
respectively. It is quite obvious that the density dependence parameter $\beta$ obtained by this method depends only on the saturation energy per nucleon $\epsilon_0$, the saturation density $\rho_{0}$ and the index $n$ of the density dependent part but not on the parameters of the M3Y interaction while the other density dependence parameter C depends on the parameters of the M3Y interaction also through the volume integral $J_v$. The energy per nucleon can be rewritten as    

\begin{equation}
 \epsilon = [3\hbar^2k_F^2/10m] - (\rho/\rho_{0}) [\hbar^2k_{F_0}^2 (1 - \beta\rho^n)]/[5m(1 - (n+1)\beta\rho_{0}^n)]
\label{seqn14}
\end{equation}
\noindent
The pressure $P$ and the energy density $\varepsilon$ of nuclear matter can be given by 

\begin{equation}
 P = \rho^2 \partial\epsilon/\partial\rho = [\rho \hbar^2k_F^2/5m] + \rho^2 J_v C [1 - (n+1) \beta\rho^n]/2,
\label{seqn15}
\end{equation} 
\noindent

\begin{equation}
 \varepsilon = \rho (\epsilon + m c^2) = \rho [(3\hbar^2k_F^2/10m) + \rho J_v C (1 - \beta\rho^n)/2 + m c^2], 
\label{seqn16}
\end{equation} 
\noindent
respectively, and thus the velocity of sound $v_s$ in standard nuclear matter is given by 

\begin{equation}
 v_s/c=\sqrt{\partial P/\partial\varepsilon}=\sqrt{[2\rho\partial\epsilon/\partial\rho-\hbar^2k_F^2/15m- J_vCn(n+1)\beta \rho^{n+1}/2]/[\epsilon + m c^2 + \rho \partial\epsilon/\partial\rho]}
\label{seqn17}
\end{equation} 
\noindent
The incompressibility $K_0$ of the spin and isospin symmetric cold infinite nuclear matter which is defined as   
  
\begin{equation}
 K_0 = k_F^2\partial^2\epsilon/\partial{k_F^2} = 9\rho^2\partial^2\epsilon/\partial\rho^2\mid_{\rho=\rho_0}
\label{seqn18}
\end{equation}
\noindent
can be theoretically obtained using eqn.(18), eqn.(6) and eqn.(9) as

\begin{equation}
 K_{0} = [-(3\hbar^2k_{F_0}^2/5m) - 9 J_v C n(n+1) \beta\rho_0^{n+1}/2]
\label{seqn19}
\end{equation} 
\noindent
Since the product $J_v C$ appears in the above equation, a cursory glance reveals that the incompressibility $K_0$ depends only upon the saturation energy per nucleon $\epsilon_0$, the saturation density $\rho_{0}$ and the index $n$ of the density dependent part of the interaction but not on the parameters of the M3Y interaction.     

\section{Calcuations of energy per nucleon, pressure, energy density and velocity of sound}
\label{section4}

      The calculations have been performed using the values of the saturation density $\rho=0.16 fm^{-3}$ and the saturation energy per nucleon $\epsilon=-16 MeV$ for the standard nuclear matter. For a fixed value of $\beta$, the parameters $\alpha$ and C can have any possible simultaneous values as determined from nuclear matter. Using the usual value of $\alpha=0.005/MeV$ for the parameter of energy dependence of the zero range pseudo-potential, the values obtained for the density dependence parameters C and $\beta$ have been presented in Table-1 for different values of the parameter $n$ along with the corresponding values of the incompressibility $K_0$. Smaller $n$ values predict softer EOS while higher values predict stiffer EOS. The form of $C(1-\beta\rho^n)$ with $n=2/3$ for the density dependence which is identical to that used for explaining the scattering and the $\alpha$, cluster radioactivity phenomena \cite{r7,r8} also agrees well with recent theoretical \cite{r10} and experimental \cite{r11} results for the nuclear incompressibility. 

\begin{table}
\caption{Compression modulus at different parameter values of n}
\begin{tabular}{ccccc}
$n$&$\beta$&$\alpha$&C & $K_{0}$      \\
 &$fm^2$&$MeV^{-1}$&          &$MeV$    \\ \hline
 1/6&1.0672&0.005&5.02&201.2 \\
 1/3&1.1934&0.005&3.05&236.2 \\
  1/2&1.3772&0.005&2.39&271.3 \\
  2/3&1.6257&0.005&2.06&306.3 \\ 
  1&2.3758 &0.005 &1.73 &376.4 \\ 
  2&9.1667 &0.005 &1.40 &586.8 \\ 
  3&41.4358 &0.005 &1.29 &797.1 \\ 
  
\end{tabular} 
\end{table}
\nopagebreak

      In Table-2 the theoretical estimates of the pressure $P$ and velocity of sound $v_s$ of standard nuclear matter have been listed as a function of nucleonic density $\epsilon$ and energy density $\varepsilon$ using the usual value of 0.005/MeV for the parameter of energy dependence, given in eqn.(3), of the zero range pseudo-potential and also the standard value of the parameter $n=2/3$. As for any other non-relativistic EOS, present EOS also suffers from superluminosity at very high densities. According to present calculations the velocity of sound becomes imaginary for $\rho\le 0.1fm^{-3}$ and exceeds the velocity of light c at $\rho \ge5.2\rho_0$ and the EOS obtained using $v_{14}+TNI$ \cite{r12} also resulted in sound velocity becoming imaginary at same nuclear density and  superluminous at about the same nuclear density. But in contrast, the incompressibility $K_0$ of infinite nuclear matter for the $v_{14}+TNI$ was 240 MeV while that by the present theoretical estimate is about 300 MeV which is in excellent agreement with the experimental value of $K_0=300\pm25$ MeV obtained from the giant monopole resonance \cite{r13} and with the the recent experimental determination of $K_0$ based upon the production of hard photons in heavy ion collisions which led to the experimental estimate of $K_0=290\pm50$ MeV \cite{r11}.

\begin{table}
\caption{Energy per nucleon $\epsilon$, Pressure $P$, Energy density $\varepsilon$ and velocity of sound $v_s$ as  functions of nuclear density $\rho$}
\begin{tabular}{cccccc}
$\rho$&$\rho/\rho_{0}$&$\epsilon$&P&$\varepsilon $&$v_s$  \\
$fm^{-3}$& &$MeV$   &$MeV fm^{-3}$&$MeV fm^{-3}$&in units of c       \\ \hline

     .01&  .6250E-01& -.7458E+00& -.1676E-01&  .9382E+01&  .0000E+00\\
     .02&  .1250E+00& -.2523E+01& -.7270E-01&  .1873E+02&  .0000E+00\\
     .03&  .1875E+00& -.4322E+01& -.1590E+00&  .2804E+02&  .0000E+00\\
     .04&  .2500E+00& -.6039E+01& -.2658E+00&  .3732E+02&  .0000E+00\\
     .05&  .3125E+00& -.7638E+01& -.3834E+00&  .4656E+02&  .0000E+00\\
     .06&  .3750E+00& -.9103E+01& -.5027E+00&  .5579E+02&  .0000E+00\\
     .07&  .4375E+00& -.1043E+02& -.6149E+00&  .6499E+02&  .0000E+00\\
     .08&  .5000E+00& -.1161E+02& -.7115E+00&  .7418E+02&  .0000E+00\\
     .09&  .5625E+00& -.1265E+02& -.7845E+00&  .8336E+02&  .0000E+00\\
     .10&  .6250E+00& -.1355E+02& -.8260E+00&  .9254E+02&  .0000E+00\\
     .11&  .6875E+00& -.1430E+02& -.8283E+00&  .1017E+03&  .4636E-01\\
     .12&  .7500E+00& -.1492E+02& -.7841E+00&  .1109E+03&  .8728E-01\\
     .13&  .8125E+00& -.1539E+02& -.6862E+00&  .1201E+03&  .1177E+00\\
     .14&  .8750E+00& -.1573E+02& -.5274E+00&  .1292E+03&  .1443E+00\\
     .15&  .9375E+00& -.1593E+02& -.3009E+00&  .1384E+03&  .1689E+00\\
     .16&  .1000E+01& -.1600E+02&  .3638E-15&  .1477E+03&  .1920E+00\\
     .17&  .1063E+01& -.1593E+02&  .3820E+00&  .1569E+03&  .2142E+00\\
     .18&  .1125E+01& -.1574E+02&  .8515E+00&  .1662E+03&  .2357E+00\\
     .19&  .1188E+01& -.1541E+02&  .1415E+01&  .1755E+03&  .2565E+00\\
     .20&  .1250E+01& -.1495E+02&  .2079E+01&  .1848E+03&  .2768E+00\\
     .21&  .1313E+01& -.1437E+02&  .2849E+01&  .1942E+03&  .2966E+00\\
     .22&  .1375E+01& -.1366E+02&  .3731E+01&  .2036E+03&  .3160E+00\\
     .23&  .1438E+01& -.1283E+02&  .4732E+01&  .2130E+03&  .3349E+00\\
     .24&  .1500E+01& -.1187E+02&  .5857E+01&  .2225E+03&  .3535E+00\\
     .25&  .1563E+01& -.1080E+02&  .7112E+01&  .2320E+03&  .3718E+00\\
     .26&  .1625E+01& -.9597E+01&  .8504E+01&  .2416E+03&  .3897E+00\\
     .27&  .1688E+01& -.8280E+01&  .1004E+02&  .2513E+03&  .4073E+00\\
     .28&  .1750E+01& -.6844E+01&  .1172E+02&  .2610E+03&  .4245E+00\\
     .29&  .1812E+01& -.5291E+01&  .1355E+02&  .2708E+03&  .4415E+00\\
     .30&  .1875E+01& -.3622E+01&  .1554E+02&  .2806E+03&  .4581E+00\\
     .31&  .1938E+01& -.1838E+01&  .1769E+02&  .2905E+03&  .4745E+00\\
     .32&  .2000E+01&  .5969E-01&  .2002E+02&  .3005E+03&  .4905E+00\\
     .33&  .2063E+01&  .2071E+01&  .2252E+02&  .3105E+03&  .5063E+00\\
     .34&  .2125E+01&  .4194E+01&  .2519E+02&  .3207E+03&  .5217E+00\\
     .35&  .2188E+01&  .6429E+01&  .2805E+02&  .3309E+03&  .5369E+00\\
     .36&  .2250E+01&  .8774E+01&  .3111E+02&  .3412E+03&  .5518E+00\\
     .37&  .2313E+01&  .1123E+02&  .3435E+02&  .3516E+03&  .5664E+00\\
     .38&  .2375E+01&  .1379E+02&  .3780E+02&  .3620E+03&  .5807E+00\\
     .39&  .2438E+01&  .1646E+02&  .4145E+02&  .3726E+03&  .5948E+00\\
     .40&  .2500E+01&  .1924E+02&  .4531E+02&  .3833E+03&  .6086E+00\\
     .41&  .2563E+01&  .2213E+02&  .4939E+02&  .3940E+03&  .6222E+00\\
     .42&  .2625E+01&  .2512E+02&  .5369E+02&  .4049E+03&  .6354E+00\\
     .43&  .2688E+01&  .2822E+02&  .5821E+02&  .4159E+03&  .6484E+00\\
     .44&  .2750E+01&  .3142E+02&  .6296E+02&  .4269E+03&  .6612E+00\\
     .45&  .2813E+01&  .3472E+02&  .6795E+02&  .4381E+03&  .6737E+00\\
     .46&  .2875E+01&  .3813E+02&  .7317E+02&  .4494E+03&  .6860E+00\\
     .47&  .2938E+01&  .4164E+02&  .7864E+02&  .4609E+03&  .6980E+00\\
     .48&  .3000E+01&  .4525E+02&  .8436E+02&  .4724E+03&  .7098E+00\\
     .49&  .3063E+01&  .4896E+02&  .9033E+02&  .4841E+03&  .7213E+00\\
     .50&  .3125E+01&  .5277E+02&  .9656E+02&  .4958E+03&  .7326E+00\\
     .51&  .3188E+01&  .5668E+02&  .1030E+03&  .5078E+03&  .7437E+00\\
     .52&  .3250E+01&  .6069E+02&  .1098E+03&  .5198E+03&  .7545E+00\\
     .53&  .3313E+01&  .6480E+02&  .1168E+03&  .5320E+03&  .7651E+00\\
     .54&  .3375E+01&  .6901E+02&  .1241E+03&  .5443E+03&  .7755E+00\\
     .55&  .3438E+01&  .7332E+02&  .1317E+03&  .5567E+03&  .7857E+00\\
     .56&  .3500E+01&  .7772E+02&  .1396E+03&  .5693E+03&  .7957E+00\\
     .57&  .3563E+01&  .8222E+02&  .1478E+03&  .5820E+03&  .8055E+00\\
     .58&  .3625E+01&  .8682E+02&  .1562E+03&  .5949E+03&  .8150E+00\\
     .59&  .3687E+01&  .9151E+02&  .1650E+03&  .6080E+03&  .8244E+00\\
     .60&  .3750E+01&  .9629E+02&  .1740E+03&  .6211E+03&  .8335E+00\\
     .61&  .3813E+01&  .1012E+03&  .1834E+03&  .6345E+03&  .8425E+00\\
     .62&  .3875E+01&  .1062E+03&  .1931E+03&  .6479E+03&  .8513E+00\\
     .63&  .3938E+01&  .1112E+03&  .2030E+03&  .6616E+03&  .8599E+00\\
     .64&  .4000E+01&  .1164E+03&  .2134E+03&  .6754E+03&  .8683E+00\\
     .65&  .4063E+01&  .1216E+03&  .2240E+03&  .6894E+03&  .8765E+00\\
     .66&  .4125E+01&  .1270E+03&  .2349E+03&  .7035E+03&  .8846E+00\\
     .67&  .4188E+01&  .1324E+03&  .2462E+03&  .7178E+03&  .8924E+00\\
     .68&  .4250E+01&  .1380E+03&  .2579E+03&  .7323E+03&  .9002E+00\\
     .69&  .4313E+01&  .1436E+03&  .2698E+03&  .7469E+03&  .9077E+00\\
     .70&  .4375E+01&  .1493E+03&  .2821E+03&  .7617E+03&  .9151E+00\\
     .71&  .4438E+01&  .1551E+03&  .2948E+03&  .7767E+03&  .9223E+00\\
     .72&  .4500E+01&  .1610E+03&  .3078E+03&  .7919E+03&  .9294E+00\\
     .73&  .4563E+01&  .1670E+03&  .3212E+03&  .8073E+03&  .9363E+00\\
     .74&  .4625E+01&  .1730E+03&  .3349E+03&  .8229E+03&  .9431E+00\\
     .75&  .4688E+01&  .1792E+03&  .3490E+03&  .8386E+03&  .9498E+00\\
     .76&  .4750E+01&  .1855E+03&  .3635E+03&  .8545E+03&  .9562E+00\\
     .77&  .4813E+01&  .1918E+03&  .3783E+03&  .8706E+03&  .9626E+00\\
     .78&  .4875E+01&  .1982E+03&  .3936E+03&  .8870E+03&  .9688E+00\\
     .79&  .4938E+01&  .2047E+03&  .4092E+03&  .9035E+03&  .9749E+00\\
     .80&  .5000E+01&  .2113E+03&  .4251E+03&  .9202E+03&  .9809E+00\\
     .81&  .5063E+01&  .2180E+03&  .4415E+03&  .9371E+03&  .9867E+00\\
     .82&  .5125E+01&  .2248E+03&  .4583E+03&  .9542E+03&  .9924E+00\\
     .83&  .5188E+01&  .2316E+03&  .4755E+03&  .9716E+03&  .9980E+00\\
     .84&  .5250E+01&  .2386E+03&  .4930E+03&  .9891E+03&  .1003E+01\\
     .85&  .5313E+01&  .2456E+03&  .5110E+03&  .1007E+04&  .1009E+01\\
     .86&  .5375E+01&  .2527E+03&  .5294E+03&  .1025E+04&  .1014E+01\\
     .87&  .5438E+01&  .2599E+03&  .5482E+03&  .1043E+04&  .1019E+01\\
     .88&  .5500E+01&  .2672E+03&  .5674E+03&  .1061E+04&  .1024E+01\\
     .89&  .5563E+01&  .2746E+03&  .5870E+03&  .1080E+04&  .1029E+01\\
     .90&  .5625E+01&  .2820E+03&  .6070E+03&  .1099E+04&  .1034E+01\\
     .91&  .5688E+01&  .2896E+03&  .6275E+03&  .1118E+04&  .1039E+01\\
     .92&  .5750E+01&  .2972E+03&  .6484E+03&  .1137E+04&  .1043E+01\\
     .93&  .5813E+01&  .3049E+03&  .6698E+03&  .1157E+04&  .1048E+01\\
     .94&  .5875E+01&  .3127E+03&  .6916E+03&  .1177E+04&  .1052E+01\\
     .95&  .5938E+01&  .3205E+03&  .7138E+03&  .1196E+04&  .1057E+01\\
     .96&  .6000E+01&  .3285E+03&  .7365E+03&  .1217E+04&  .1061E+01\\
     .97&  .6063E+01&  .3365E+03&  .7596E+03&  .1237E+04&  .1065E+01\\
     .98&  .6125E+01&  .3446E+03&  .7832E+03&  .1258E+04&  .1069E+01\\
     .99&  .6188E+01&  .3528E+03&  .8072E+03&  .1279E+04&  .1073E+01\\
    1.00&  .6250E+01&  .3611E+03&  .8317E+03&  .1300E+04&  .1077E+01\\

\end{tabular} 
\end{table}
\nopagebreak

      In Fig.-1 plots of the theoretical estimates of the velocity of sound and the energy per nucleon $\epsilon$ of the standard nuclear matter have been shown as a function of the normalized nuclear density $\rho/\rho_0$ using $n=2/3$. The continuous line represents the energy per nucleon in $MeV$ and the dash-dotted line represents the velocity of sound in units of $10^{-2}c$ for the standard nuclear matter. Fig.-2 presents the plots of the pressure $P$ and the energy density $\varepsilon$ of the standard nuclear matter as a function of nucleonic density $\rho$ using $n=2/3$. The continuous line represents the pressure in $MeV fm^{-3}$ and the dash-dotted line represents energy density in the same units for the standard nuclear matter. 

\section{Results and discussions}
\label{section5}

      The theoretical estimate $K_0$ of the incompressibility of spin-isospin symmmetric infinite nuclear matter  obtained from present approach using DDM3Y is about $300 MeV$. The theoretical estimate of $K_0$ from the refractive $\alpha$-nucleus scattering is about 240 MeV-270 MeV \cite{r14,r15} and that by infinite nuclear matter model (INM) \cite{r10} claims a well defined and stable value of $K_0=288\pm20$ MeV and present theoretical estimate is in reasonably close agreement with the value obtained by INM which rules out any values lower than 200 MeV. Using the value of the density dependence parameter $\beta=1.6257 fm^2$ corresponding to the standard value of the parameter $n=2/3$ along with the nucleonic density of $0.16 fm^{-3}$, the value obtained for the nuclear mean free path $\lambda$ is about $4 fm$ which is in excellent agreement \cite{r16} with other theoretical estimates. 
 
      There is, however, some uncertaity about the saturation energy per nucleon $\epsilon=-16 MeV$ for the standard nuclear matter. A five parameter least square fit to the Bethe-Weizs\"acker mass formula for theoretical binding energies $B(A,Z)_{Th} = a_vA-a_sA^{2/3}-a_cZ(Z-1)/A^{1/3}-a_{sym}(A-2Z)^2/A+\delta$ where $\delta=a_pA^{-1/2}$ for even N-even Z, $=-a_pA^{-1/2}$ for odd N-odd Z, $=0$ for odd A, has been achieved by $\sigma^2 =  (1/N) \sum [ B(A,Z)_{Th}  - B(A,Z)_{Ex} ]^2$ minimization, where $B(A,Z)_{Ex}$ are the experimental binding energies from the recent Audi-Wapstra mass table \cite{r17} and the summation runs over all the N=3179 nuclei listed in the table which yields the new parameter values as $a_v=15.25 \pm 0.02 ~MeV, ~a_s=16.25 \pm 0.06 ~MeV, ~a_c=0.6877 \pm 0.0014 ~MeV, ~a_{sym}=22.18 \pm 0.05 ~MeV$ and $a_p=10.07 \pm 0.85 ~MeV$. Identifying $a_v$ as saturation energy per nucleon for standard nuclear matter and hence using the saturation energy per nucleon $\epsilon=-15.25 MeV$, the saturation density $\rho=0.16 fm^{-3}$, $n=2/3$ and $\alpha= 0.005/MeV$ the values obtained for the energy dependence parameters and nuclear incompressibility are $\beta=1.6149 fm^2$, C=   2.01 and $K_{0}= 295.1 MeV$ respectively.  

\section{Summary and conclusions}
\label{section6}
 
      In summary, we conclude that the present EOS which is obtained using the M3Y effective NN interaction, which was derived by fitting its matrix elements in an oscillator basis to those elements of the G-matrix obtained with the Reid-Elliot soft-core NN interaction, has a profound theoretical standing. The form $C(1-\beta\rho^{2/3})$ of the density dependence is also physical and provides excellent results for nuclear matter, $\alpha$ and heavy ion scatterings as well as $\alpha$ and heavy cluster radioactivities. The value obtained for the nuclear mean free path is in excellent agreement \cite{r16} with other theoretical estimates. The present theoretical estimate of nuclear incompressibility is in reasonably close agreement with other theoretical estimates obtained by INM \cite{r10} model, using the Seyler-Blanchard interaction \cite{r18} or the relativistic Brueckner-Hartree-Fock theory \cite{r19}. This value is also in excellent agreement with the experimental estimates from GMR \cite{r13} as well as determination based upon the production of hard photons in heavy ion collisions \cite{r11}. Thus, a unified description for nuclear scattering, radioactivity and EOS for standard nuclear matter has been achieved using a realistic effective interaction along with a physical density dependence.

\begin{figure}[h]
\eject\centerline{\epsfig{file=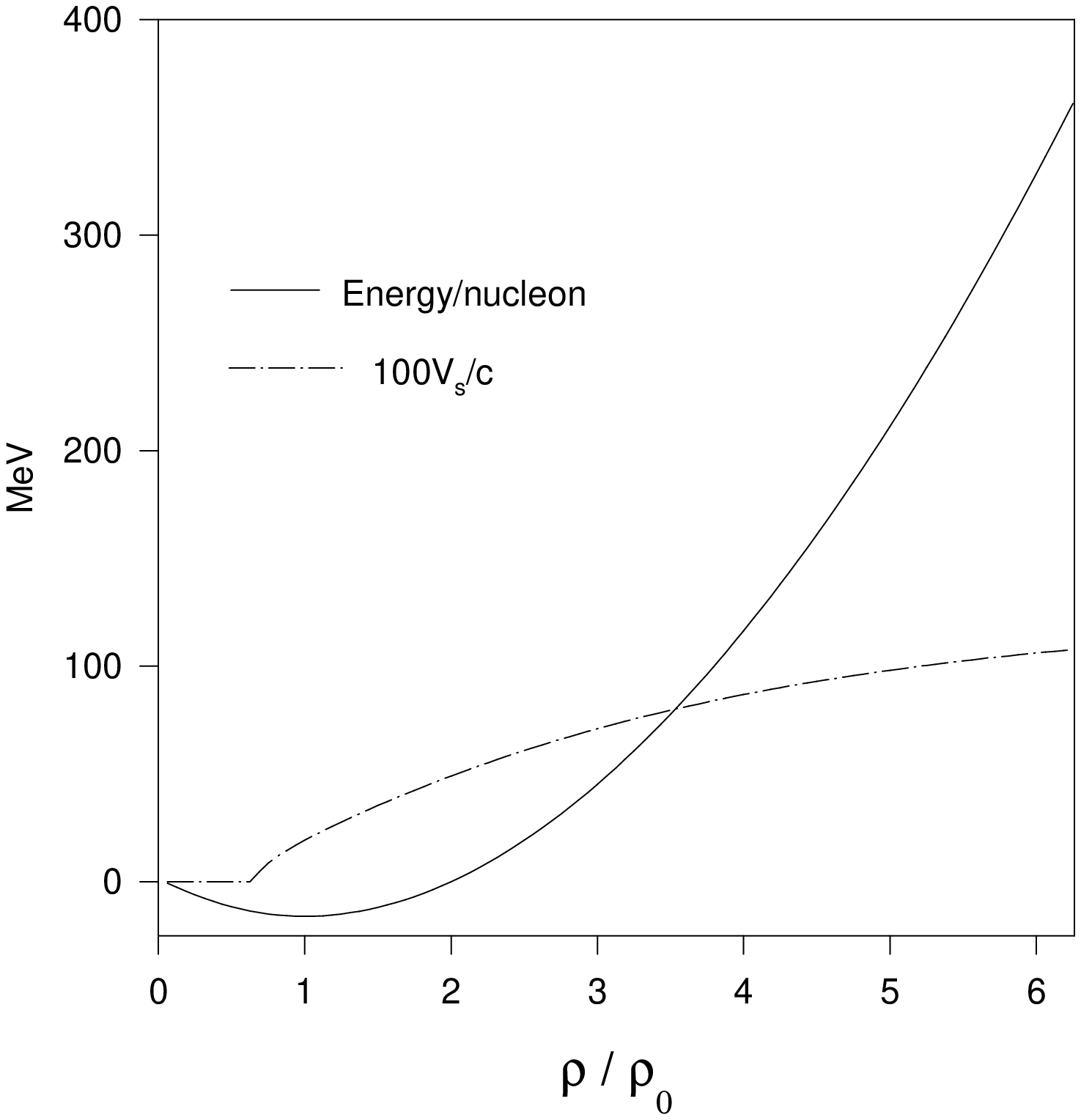,height=15cm,width=10cm}}
\caption
{The energy per nucleon $\epsilon$ and the velocity of sound $v_s$ in nuclear matter as a function of $\rho/\rho_0$. The continuous line represents the energy per nucleon in $MeV$ and the dash-dotted line represents the velocity of sound in units of $10^{-2}c$ for the standard nuclear matter using $n=2/3$.}
\label{fig1}
\end{figure}

\begin{figure}[h]
\eject\centerline{\epsfig{file=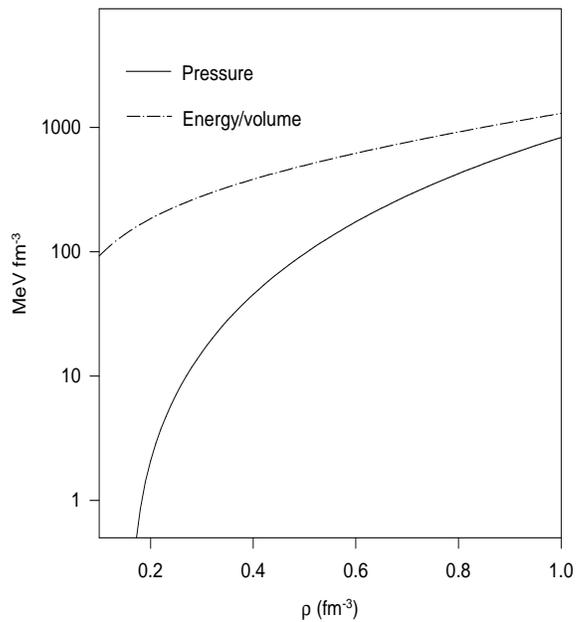,height=15cm,width=10cm}}
\caption
{The pressure $P$ and energy density $\varepsilon$ of nuclear matter as a function of nuleonic desity $\rho$. The continuous line represents the pressure in $MeV fm^{-3}$ and the dash-dotted line represents energy density in the same units for the standard nuclear matter using $n=2/3$.}
\label{fig2}
\end{figure}

\end{document}